# MAGNETIC MODULATION IN MECHANICAL ALLOYED $Cr_{1.4}Fe_{0.6}O_3$ OXIDE

R.N. Bhowmik, Nrisimha Murty.M and Sekhar Srinadhu. E

Department of Physics, Pondicherry University, R. Venkataraman Nagar,

Kalapet, Pondicherry-605014, India.

E-mail of corresponding author (RNB): rnbhowmik.phy@pondiuni.edu.in

**ABSTRACT:**

We have synthesized $Cr_{1.4}Fe_{0.6}O_3$ compound through mechanical alloying of $Cr_2O_3$ and $\alpha$-$Fe_2O_3$ powders and subsequent thermal annealing. The XRD spectrum, SEM picture and microanalysis of EDAX spectrum have been used to understand the structural evolution in the alloyed compound. The alloyed samples are matching to rhombohedral structure with R3C space group. The observation of a modulated magnetic order confirmed a systematic diffusion of Fe atoms into the Cr sites of lattice structure. A field induced magnetic behaviour is seen in the field dependence of magnetization data of the annealed samples. The behaviour is significantly different from the mechanical alloyed samples. The experimental results provided the indications of considering the present material as a potential candidate for opto-electronic applications.

## I. INTRODUCTION:

In recent years extensive research efforts have been given to the formation of new magnetic oxides, considering their potential applications in the field of micro-electronics and multifunctional devices [1, 2]. In order to search for new class of materials, it has been found that solid solution of mixed metal oxides might be a potential candidate, having advantage of easy alloying due to ionic radius of same order [3, 4]. $Fe_{2-x}Cr_xO_3$ is one such series compound, which can be formed through the alloying of $\alpha$-$Fe_2O_3$ and $Cr_2O_3$ oxides [5, 6]. The interesting points are that both $\alpha$-$Fe_2O_3$ and $Cr_2O_3$ stabilize into rhombohedral crystal structure with space group R3C [4, 7] and both are antiferromagnetic insulator [8]. $Cr_2O_3$ oxide is a well known compound for the prediction and experimental observation of large magneto-electric (ME) effect [9, 10]. The ME effect has been found to be extremely small in symmetric non-magnetic insulators [11]. The recent work by Matteo et al. [12] suggested a possible change of magnetic space group symmetry in $Cr_2O_3$ for the exhibition of large magnet-electric effect. However, the low magnetic moment and low antiferromagnetic ordering temperature ($T_N$) of single phase $Cr_2O_3$ (~310 K) is not suitable for exhibiting large magneto-electric effect, as well as for applications [13, 14]. This triggered the alloying of $Fe_{2-x}Cr_xO_3$ series by mixing a suitable amount of $Fe_2O_3$ ($T_N$ ~950 K) [4, 8] in $Cr_2O_3$ oxide. The study of this series remained attractive for the last few decades for its promising applications in the field of opto-electronic materials. The immediate effect is that photo-conductivity in $Fe_{2-x}Cr_xO_3$ series has shown enhancement with the increase of Cr content [2, 15]. In order to understand the modified photo-conductivity, as well as magneto-electric effect, one needs to realize the correlation between crystal structure and magnetic ordering in the compound. The correlation between its crystal structure and magnetism would also be relevant to gain the properties of Cr-Fe interface [2].

The antiferromagnetic ordering of spins in both α-$Fe_2O_3$ and $Cr_2O_3$ has been explained in terms of super exchange interactions (Cr-O-Cr, Fe-O-Fe) [4, 8]. Neutron diffraction study attributed the drastic variation of $T_N$ (~310 K for $Cr_2O_3$ and ~950 K for $Fe_2O_3$) to a different kind of magnetic structure, although α-$Fe_2O_3$ and $Cr_2O_3$ have shown identical crystal structure. Earlier reports [4, 7, 8] suggested that the spin moments of $Cr^{3+}$ ($3d^3$) ions in $Cr_2O_3$ are arranged in + - + - (+ ≡ up spin, - ≡ down spin) sequence along the [111] axis, whereas the spins of $Fe^{3+}$ ($3d^5$) ions in α-$Fe_2O_3$ are ordered in the + - - + sequence. Hence, it is expected that the substitution of $Fe^{3+}$ ($3d^5$) by $Cr^{3+}$ ($3d^3$) in $Fe_{2-x}Cr_xO_3$ solid solution would modify the sequence of spins ordering, as well as the nature of superexchange interactions. Form the phase diagram of $Fe_{2-x}Cr_xO_3$ [4] one could estimate an equal probability of Fe-O-Fe, Fe-O-Cr, Cr-O-Cr superexchange interactions near to x=1 and no significant change in the ordering of spins. This resulted in a slow variation of $T_N$ with Cr content in the region 0.90<x≤1.52 [4]. However, the nature of spin ordering may not follow the conventional structure due to the change of magnetic space group symmetry. The modulated (perturbed) local magnetic order could be expected during the diffusion of Fe atoms into Cr sites. To our knowledge, the magnetic ordering of $Fe_{2-x}Cr_xO_3$ compound is not clear till date. Attempts were made to understand the structural and magnetic properties of $Fe_{2-x}Cr_xO_3$ compound by reducing the particle size in nanometer range using various chemical routes [16, 17]. However, enough attention was not given in the reported works to realize such important issues related to the effect of modulated spin structure on the properties of $Fe_{2-x}Cr_xO_3$ compound.

In this work, we have synthesized $Cr_{1.4}Fe_{0.6}O_3$ compound using the novel technique of Mechanical alloying [18]. We have investigated the associated structural and magnetic evolution in different states of mechanical alloying and annealing of the samples. We also attempted to identify the nature of magnetic order in $Cr_{1.4}Fe_{0.6}O_3$, whether antiferromagnet or ferromagnet or the mixed properties of ferromagnet and antiferromagnet.

## II. EXPERIMENTAL

### A. Sample preparation

The stoichiometric amounts of high purity $\alpha$-$Fe_2O_3$ and $Cr_2O_3$ were mixed for the preparation of $Cr_{1.4}Fe_{0.6}O_3$ compound. The initial colours of $\alpha$-$Fe_2O_3$ and $Cr_2O_3$ are red and green, respectively. The mixture of $Fe_2O_3$ and $Cr_2O_3$ was ground using agate mortar and pestle for nearly two hours in atmospheric conditions. The ground powder was mechanical alloyed using Fritsch planetary mono mill (pulverisette 7). The material and balls (combination of 10mm Silicon Nitride and 5mm Tungsten Carbide) mass ratio was maintained to 1:7. The mechanical alloying was carried out in atmospheric condition up to 84 hours in a silicon nitride ($Si_3N_4$) bowl with rotational speed 300 rpm. The non-magnetic balls and bowl (Silicon Nitride and Tungsten Carbide) were selected to avoid the magnetic contamination during the milling process, if it happened at al. The milling process was intermediately stopped to monitor the uniform alloying of the mixture and to minimize the local heat generation that might occur during continuous milling. A small quantity of alloyed sample after 24 hours and 48 hours was taken out to check the structural phase evolution. The alloyed samples with different milling hours were made into pellets. The pellets of 84 hours milled sample were placed in alumina crucibles and annealed at 700$^o$C in atmospheric conditions. After annealing for 1 hour, 3 hours and 17 hours, individual pellet was directly air quenched to room temperature. We denoted the mechanical alloyed samples as MA**h**, where **h** = 0, 24, 48 and 84 for alloying time 0, 24 hours, 48 hours and 84 hours, respectively. The samples annealed at $700^0$C are denoted as SN**t**, where **t** = 1, 3 and 17 for annealing time 1 hour, 3 hours and 17 hours, respectively.

**B. Sample characterization and measurements**

The crystalline phase of alloyed and annealed samples was characterized by recording the X-ray Diffraction spectra in the $2\theta$ range 10-$90^0$ with step size $0.01^0$. The Cu-K$_\alpha$ radiation from the X-ray Diffractometer (model: X pert Panalytical) were employed to record the room temperature spectrum of each sample. The scanning electron microscopic (SEM) picture of the samples was taken using HITACHI S-3400N model. Elemental composition of the samples was obtained from the energy dispersive analysis of x-ray fluorescence (EDAXF) spectrum. The magnetic properties of the samples were studied by the measurement of magnetization as a function of temperature and magnetic field using vibrating sample magnetometer (Model: Lakeshore 7400). The temperature (300 K-900 K) dependence of magnetization was measured by attaching a high temperature oven to the vibrating sample magnetometer. The temperature dependence of magnetization was carried out at 1 kOe magnetic field by increasing the temperature from 300 K to 900 K (ZFC mode) and reversing back the temperature to 300 K in the presence of same applied field 1 kOe (FC mode). It should be noted that the ZFC mode followed here is slightly different from the conventional zero field cooling (ZFC) measurement, where the sample is first cooled without applying magnetic field from the temperature greater than $T_C$ to the temperature lower than $T_C$ and magnetization measurement in the presence of magnetic field starts with the increase of temperature. The field dependence of magnetization at 300 K was measured within ±15 kOe.

## III. RESULTS AND DISCUSSION

### (a) Structural properties

Fig.1 shows the XRD spectrum of mechanical alloyed and subsequent annealed samples. The MA0 sample (before milling) shows (Fig. 1a) a mixed pattern of $Fe_2O_3$ and $Cr_2O_3$ phases. The peaks of $Fe_2O_3$ and $Cr_2O_3$ phases are well separated before mechanical alloying. As the alloying process continued by increasing the milling time, the individual peaks of $Fe_2O_3$ and $Cr_2O_3$ phases are merging to each other. The intensity of peaks is decreasing as the milling time increases up to 84 hours. At the same time broadness and symmetry in shape of the peaks are also increasing. Subsequent annealing of MA84 sample at $700^0C$, again, increases the peak intensity, along with increasing sharpness, with the increase of annealing time up to 17 hours. The undergoing phase evolution during the mechanical alloying and annealing process is clearly demonstrated in Fig. 1(b) for (104) and (110) peak positions. In the (MA0) mixed sample, the peaks position of $Cr_2O_3$ sample are at higher $2\theta$ value with respect to $Fe_2O_3$ sample. The notable feature is that peaks of $Fe_2O_3$ sample disappear with the increase of milling time and the XRD pattern of alloyed compound is tending to achieve the character of $Cr_2O_3$. This indicates the diffusion of Fe atoms into the lattice sites of Cr atoms to form a solid solution of $Cr_{1.4}Fe_{0.6}O_3$.

Now, we understand the grain size refinement process. We have determined the grain size of alloyed samples from the prominent (012), (104), (110) and (116) XRD peaks using Debye-Scherrer formula: $\langle d \rangle = \frac{0.089 * 180 * \lambda}{3.14 * \omega * \cos\theta_c} nm$, where $2\theta_C$ is the position of peak center, $\lambda$ is the wavelength of X-ray radiation (1.54056 Å), $\omega$ is the full width at the half maximum of peak height (in degrees). The average grain size of the alloyed compound is found to be in nanometer range (shown in Table I), which suggested the nanocrystalline structure of the material. The grain size of the alloyed compound slowly decreases with the increase of milling time (i.e., <d> ~ 26 nm for MA24, 22 nm for MA48, and 15 nm for MA84 sample). The thermal annealing of MA84 (lowest grain size) sample at $700^0C$ activates the increase of grain size. The lattice parameter of the samples (shown in Table I) was calculated by matching the XRD peaks into rhombohedral structure with R3C space group. The lattice parameters (*a* and *c*) and cell volume of the alloyed compound were increasing with the milling time. But, the lattice parameters of the annealed samples showed decreasing values with the increase of annealing time at $700^0C$. We understand the variation of cell parameters from the fact that lattice parameters of bulk $Cr_2O_3$ sample (*a* = 4.9510 Å, *c* =13.5996Å, cell volume = 288.69Å$^3$) is smaller than that for bulk α-$Fe_2O_3$ sample (*a* = 5.0386Å, *c* =13.7498Å, cell volume = 302.3Å$^3$). Our calculated lattice parameters for both bulk samples are consistent with literature values [4, 5]. The continuous increase of lattice parameters and cell volume with milling time indicates the kinetic diffusion of larger size (0.67 Å) $Fe^{3+}$ ions into the boundary of smaller size (0.65 Å) $Cr^{3+}$ ions [2, 4, 19]. On the other hand, the decrease of cell parameters with the increase of annealing time indicated that the alloying process is, still, continued to attain the crystal structure of $Cr_2O_3$ dominated phase, because atomic ratio of Cr and Fe is 7:3. Here, we would like to point out that thermal heating is playing a major role for obtaining the $Cr_2O_3$ dominated phase. We believe that the variation of cell parameters is not significantly affected by the nano-sized grains in comparison with the undergoing change during alloying and annealing process. The cell parameters and grain size of the MA0 sample were not calculated, as the pattern is not due to single phase XRD spectrum.

Now, we study the surface morphology (SEM picture) and chemical composition (EDAX spectrum) for selected (MA0, MA48, MA84 and SN17) samples. The selected zone (~80 μm x 80 μm) of the samples, along with SEM picture and EDAXF spectrum, are shown in Fig. 2. The SEM picture of MA0 sample (Fig. 2a) suggests the heterogeneous mixed character of α-$Fe_2O_3$ and $Cr_2O_3$ particles. The mechanical alloying between α-$Fe_2O_3$ and $Cr_2O_3$ particles can be understood from the SEM picture of MA84 sample (Fig. 2b), where nanoparticles are distributed with a little bit agglomeration. A better uniformity in the particle size distribution is observed from the SEM picture (Fig. 2c) of thermal activated (SN17) sample. The elemental composition of the samples is determined from the point and shoots microanalysis at 10 selected points over a selected zone. The elements are found to be Cr, Fe and O. The atomic percentage of the elements is estimated from the EDAXF spectrum of the samples and Fig. 2d represents the spectrum for MA84 sample. The composition of Cr and Fe varies from 0.33 to 1.34 and 1.64 to 0.57, respectively, with respect to O composition 3 in MA0 sample. This suggests a random mixing of $Fe_2O_3$ and $Cr_2O_3$ oxides (i.e., an inhomogeneous distribution of elements) in MA0 sample. We obtained the atomic % of Fe (12.12), Cr (27.88) and O (60) (i.e., elemental composition $Cr_{1.38}Fe_{0.6}O_{2.97}$) for MA84 sample; atomic % of Fe (11.96), Cr (27.06), O (60.17) (i.e., elemental composition $Cr_{1.36}Fe_{0.6}O_{3.01}$) for SN17 sample. Therefore, elemental composition of the alloyed as well as annealed samples is close to the expected composition ($Cr_{1.4}Fe_{0.6}O_3$). The distribution of Cr, Fe and O atoms in the alloyed samples are cross checked from elemental mapping over a selected zone, as shown in Fig. 3a for MA48 sample. The particle size (~ 500 nm) of MA48 sample estimated from the SEM picture (Fig. 3b) is much larger than that obtained from XRD data (~ 22 nm). This represents the fact that SEM is not the suitable tool to estimate the particle size, as the scale (micrometer range) of SEM data is larger in comparison with the particle size in nanometer range. This means the SEM picture indicates the size of multi-grained particles. The mapping (Fig. 3c-e) suggests a uniform distribution of Cr, Fe, O atoms over the zone. Achieving some basic knowledge of the structural properties (i.e., size, shape, composition, and crystal structure) of the samples, we have attempted below to understand the magnetic properties of the samples.

**(b) Magnetic Properties**

Fig. 4 shows the temperature dependence of ZFC and FC magnetization curve of different samples. The zero field cooled magnetization (MZFC), measured at 1 kOe field, does not change significantly (plateau like behaviour) in the temperature range 300 K to 600 K. The MZFC increases rapidly above 650 K to show a peak at about $T_m \sim 810$ K and then, MZFC rapidly decreases to indicate weak temperature dependence above 870 K. On the other hand, field cooled magnetization (MFC) at 1 kOe separates out from MZFC (i.e., onset of magnetic irreversibility) below the irreversibility temperature $T_{irr} \sim 860$ K. The separation between MFC and MZFC continued to increase with the decrease of temperature down to 300 K. The above features of MZFC and MFC are seen in the bulk $\alpha$-$Fe_2O_3$ sample, as well as in the alloyed $Cr_{1.4}Fe_{0.6}O_3$ samples, except some typical changes in the shape of the curves. The continuous increase of MFC with down curvature is observed in both bulk $\alpha$-$Fe_2O_3$ and alloyed samples (e.g., MA48 and MA84). Interestingly, the shape of MFC curves of the annealed samples (e.g., SN3) has transformed into up curvature as an effect of thermal heating. The magnetization of bulk $\alpha$-$Fe_2O_3$ sample decreases in the alloyed samples. The decrease is, further, accelerated by subsequent annealing of MA84 sample at $700^0$C. The decrease of magnetization is also reflected in the suppression of peak magnetization at $T_m$, when the sample changes from bulk $\alpha$-$Fe_2O_3$ to SN17. We noted that the position of MZFC maximum at $T_m$ is highly sensitive to the magnitude of measurement field. This is confirmed from the decrease of $T_m \sim 810$ K (at 1 kOe) to 750 K (at 2 kOe) in MA48 sample (Fig. 4b). The paramagnetic to canted ferromagnetic ordering temperature $T_N \sim 950$ K of bulk $\alpha$-$Fe_2O_3$ (hematite) sample [reported in Ref. 4] is higher than $T_m \sim 810$ K at 1 kOe (in the present work). The field dependence of $T_m$ ($\sim 940$ K at 100 Oe and $\sim 845$ K at 200 Oe) has already reported in bulk $\alpha$-$Fe_2O_3$ (hematite) sample [20, 21]. The present work showed that $T_{irr}$ ($\sim 860$ K) is higher than $T_m$ ($\sim 810$ K) and $T_m$ decreases with the increase of measurement field. Based on these facts, we suggest that $T_m$ does not represent a paramagnetic to canted ferromagnetic ordering temperature of the samples. Rather, $T_m$ can be attributed as the temperature below which magnetic domains are blocked along the local anisotropy axes, where anisotropy energy might be higher than the measurement field [22, 23, 24]. The interesting point is that magnetic phase of $\alpha$-$Fe_2O_3$ is, still, dominating in the alloyed compound, of course with diminished magnitude with milling time. This can be attributed to the presence of larger atomic moment of $Fe^{3+}$ ions ($\sim 5\mu_B$) in comparison with $Cr^{3+}$ ions ($\sim 3\mu_B$). The effect of $\alpha$-$Fe_2O_3$ phase is becoming weak in the annealed samples, as seen from the weak magnetic irreversibility and upward curvature in the MFC curve of SN3 sample. The results clearly showed that $Cr_2O_3$ dominated typical antiferromagnetic phase in the alloyed compound is yet to achieve at $700^0$C. The diffusion of Fe atoms into the Cr lattices either still continues, although the XRD data indicated the stabilization of alloyed $Cr_{1.4}Fe_{0.6}O_3$ close to the pattern of $Cr_2O_3$ sample. The appearance of a different kind of magnetic behaviour in the annealed samples may also be attributed to a modulated magnetic ordering at the interface of nano-sized grains.

We, now, investigate the existence of possible modulated magnetic phase in the samples from the field dependence of dc magnetization. The magnetization (M) with the variation of field (H) at room temperature is shown in Fig 5. The M(H) data of bulk $Fe_2O_3$ and $Cr_2O_3$ are shown in Fig. 5a to compare the magnetic change in alloyed compound. Bulk $Cr_2O_3$ sample does not show any noticeable hysteresis loop, whereas $Fe_2O_3$ sample has shown a well defined loop. The mixture of $Fe_2O_3$ and $Cr_2O_3$, i.e., MA0 sample (Fig. 5b), shows a clear loop, but the magnitude of loop of MA0 sample is less in comparison with $Fe_2O_3$. This is due to the effect of $Cr_2O_3$ coexisting with $Fe_2O_3$. As the alloying process continued by increasing the milling time, the loop area gradually decreases, but noticeable, as shown for MA48 sample in the inset of Fig.5c. It is interesting to note that the nature of M(H) curves of annealed samples (Fig. 5e-f) drastically differ in comparison with the mechanical alloyed samples. The extrapolation of high field M (positive H side) data (shown by dotted line in Fig. 5f) intersects the M axis to negative value and the negative magnetization confirmed the field induced magnetic behaviour in the annealed samples. We would like to mention that neither bulk $Fe_2O_3$ nor bulk $Cr_2O_3$ sample exhibited such induced magnetization. Recently, similar field induced magnetization has been observed in the $Cr_2O_3$ nanoparticles [25], originating from the competitive spin ordering at the core-shell interface [26]. Viewing the absence of such field induced magnetic ordering in mechanical alloyed (with out heat treatment) samples, although grain size is in nanometer range, we believe that the field induced magnetic behaviour in our annealed samples is not solely due to particle size effect, but can be attributed to a modulated magnetic ordering due to diffusion of two different magnetic elements [27]. The induced magnetic behaviour in annealed samples is also reflected in the field dependence of differential magnetization (dM/dH) curves (Fig. 6). The differential susceptibility (dM/dH) increases to exhibit a maximum for mechanical alloyed samples when the applied field approaches to zero value, whereas dM/dH exhibits a minimum near to the zero field value for the annealed samples. The observation of induced magnetism confirms an experimental evidence of modulated spin structure in this compound, which was suggested from Neutron diffraction study [8]. To get the insight of modulated magnetic ordering, we analyze the M(H) data in terms of a general power series of H, i.e., the combination of liner and non-linear components in the equation

$$M(H) = \chi_1 H + \chi_2 H^2 + \chi_3 H^3 + \text{(rest of the terms)} \qquad (1)$$

In the above equation, we have assumed that magnetic susceptibility is not a scalar quantity, rather a tensor with directional property in the material. The coefficients $\chi_1$ and $\chi_3$ are assumed to be symmetric with field (H), where as $\chi_2$ is antisymmetry to preserve the vector nature of magnetization (M). The validity of equation (1) is tested from Fig. 7. The M vs. $H^2$ data (in Fig. 7a) indicate that magnetization of all the samples are symmetric about the $H^2$ axis, irrespective of alloyed or annealed samples, and the change of magnetic direction is taken care by $\chi_2$, i.e., $\chi_2(-H) = -\chi_2(H)$. It is really interesting to note that M vs. $H^3$ plot (Fig. 7b) represents a ferromagnetic like appearance, irrespective of the samples and confirms the symmetric nature of $\chi_3$, where the change of magnetization direction is taken care by H. The investigation of non-linear components (specially the second-harmonic generation) in antiferromagnetic material may be one of the important technological aspects of the present material.

Now, we look at the magnetic parameters. The remanent magnetization ($M_R$) and coercive field ($H_C$) of the samples were calculated using M(H) data. The variation of $M_R$ and

$H_C$ of the alloyed compound with milling time and annealing time is shown in Fig. 8. We noted that both $H_C$ (~1620 Oe) and $M_R$ (~172 memu/g) of $\alpha$-Fe$_2$O$_3$ sample are reduced by nearly 2.5 times in MA0 sample ($H_C$ ~700 Oe and $M_R$ ~66.3 memu/g), although Cr$_2$O$_3$ is not showing any hysteresis loop. This feature suggests that properties of magnetic particles, irrespective of bulk (present work) or nanosize (reported work [28]), are strongly influenced by the surrounding matrix. The $H_C$ of MA0 sample is continued to decrease with milling time (e.g., $H_C$ ~120 Oe for MA24). On the other hand, $H_C$ has shown a significant increase by annealing MA84 sample at 700$^0$C up to 3 hours, followed by a slight decrease at 17 hours annealed sample (SN17). The $M_R$ of MA0 sample also decreases to attain an almost constant value ~3.4 memu/g for MA24 and MA48 samples, followed by a slow increase (~4.4 memu/g) for MA84 sample and continued for (~5.3 memu/g) SN1 sample. The $M_R$ value, again, starts to decrease at higher annealing time, e.g. ~ 3.9 memu/g for SN17 sample. The decrease of both remanent magnetization and coercivity at higher annealing time indicates that the alloyed compound is approaching to the magnetic state dominated by non-hysteretic Cr$_2$O$_3$ sample or a different kind of magnetic ordering that is already indicated earlier.

## IV. CONCLUSIONS

The magnetic properties of mechanical alloyed Cr$_{1.4}$Fe$_{0.6}$O$_3$ compound strongly depends on the structural change associated with the variation of milling time and annealing temperature. The variation of cell parameters is attributed to the effect of diffusion of Fe$^{3+}$ ions into the boundary of Cr$^{3+}$ ions. The structural analysis, using XRD and SEM with EDAX spectrum, may suggest the completion of alloy formation during milling process, but the magnetic behaviour confirmed that the alloying process is, still, continued at 700$^0$C to approach the structure dominated by $Cr_2O_3$ phase. Experimental results suggested that magnetic ordering of mechanical alloyed $Cr_{1.4}Fe_{0.6}O_3$ compound, annealed at $700^0C$, belongs to neither a canted ferromagnetic ordering of $\alpha$-$Fe_2O_3$ nor a typical antiferromagnetic ordering of $Cr_2O_3$. Rather, a field induced magnetic order appeared in the annealed samples, which we attribute to the appearance of a modulated spin structure in the compound. This experimental work may instigate subsequent investigations on the non-linear magneto-opto-electronic properties of similar materials.

**Acknowledgment:** We thank to CIF, Pondicherry University for providing experimental facilities. We also thank to FIST Program in the department of Physics for providing XRD measurement facilities. RNB also thanks UGC for providing financial support (F.No. 33-5/2007(SR)).

**Table. I:** Grain size and cell parameters of the samples are calculated using XRD data. The parameters are not calculated for MA sample, due to the coexistence of two XRD patterns.

| Sample | Grain size (nm) | Lattice parameter $a$(Å) | Lattice parameter $c$(Å) | Volume (Å)$^3$ |
| --- | --- | --- | --- | --- |
| $\alpha$-$Fe_2O_3$ | -- | 5.0386 ± 0.0014 | 13.7498 ± 0.0002 | 302.30 ± 0.0840 |
| MA0 | -- | -- | -- | -- |
| MA24 | 26 | 4.9521 ± 0.0021 | 13.6063 ± 0.0004 | 288.96 ± 0.1221 |
| MA48 | 22 | 4.9535 ± 0.0016 | 13.6041 ± 0.0002 | 289.08 ± 0.0923 |
| MA84 | 15 | 4.9546 ± 0.0013 | 13.6082 ± 0.0003 | 289.29 ± 0.0781 |
| SN1 | 19 | 4.9526 ± 0.0019 | 13.6142 ± 0.0002 | 289.19 ± 0.1090 |
| SN3 | 21 | 4.9520 ± 0.0018 | 13.6084 ± 0.0002 | 288.99 ± 0.1099 |
| SN17 | 22 | 4.9512 ± 0.0012 | 13.6070 ± 0.0002 | 288.87 ± 0.0997 |
| $Cr_2O_3$ | -- | 4.9510 ± 0.0011 | 13.5996 ± 0.0001 | 288.69 ± 0.0667 |

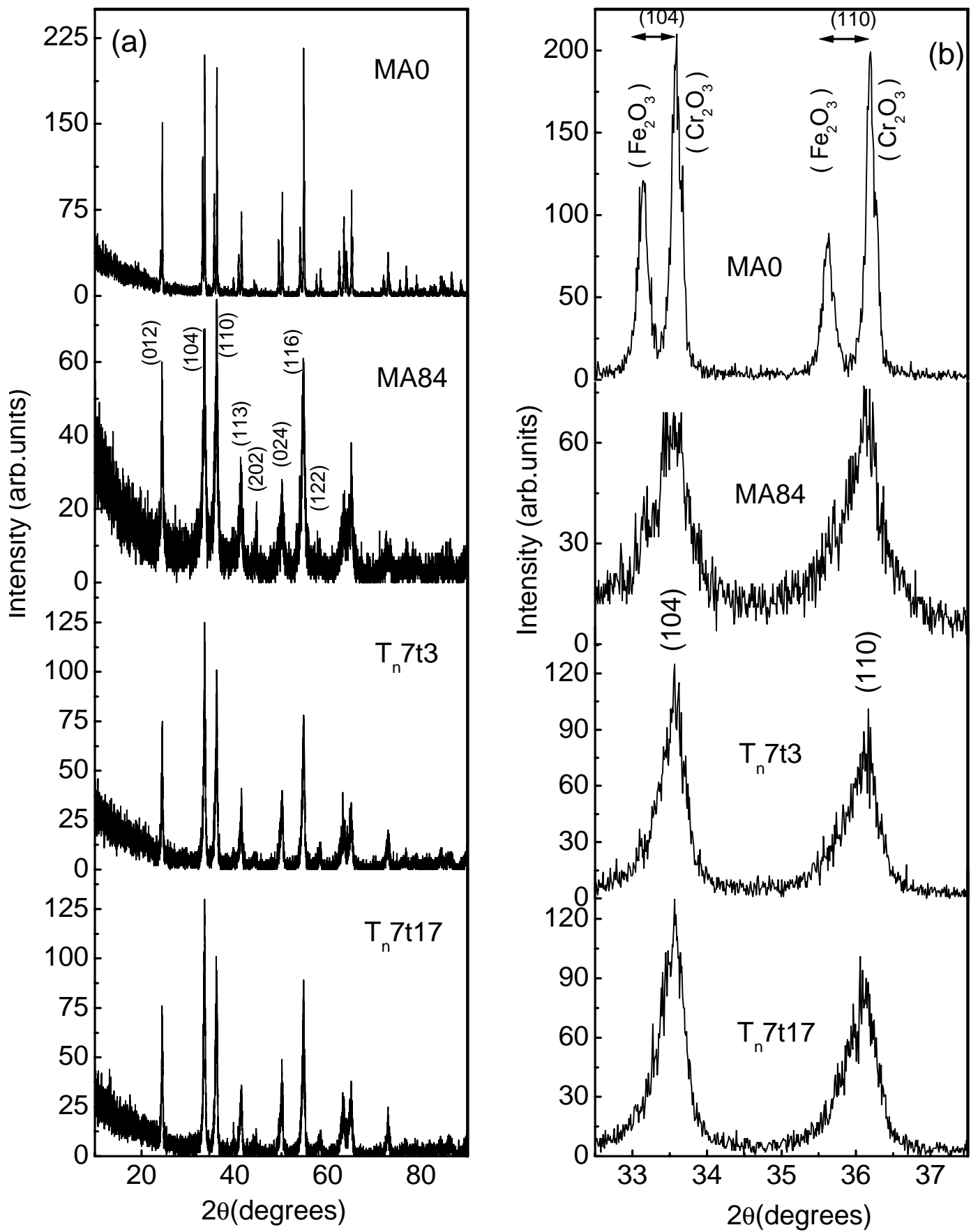

Fig. 1. (a) XRD spectrum of alloyed samples with milling time and annealing time. The (104) and (110) XRD peaks of the samples are shown in (b).

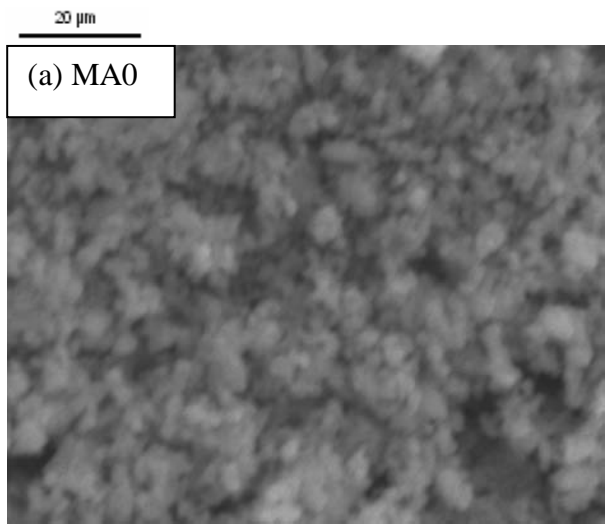
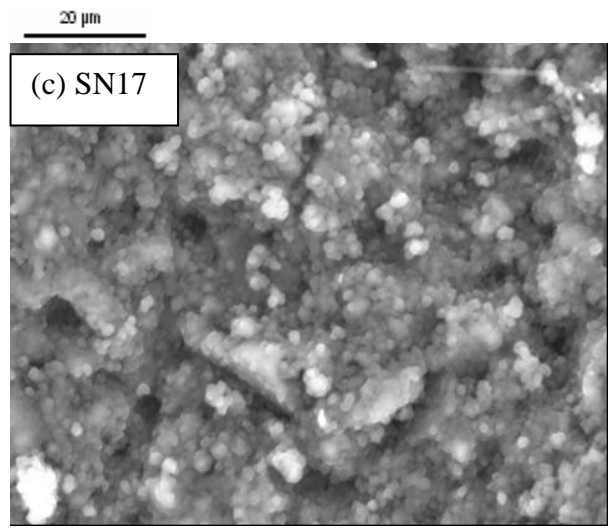
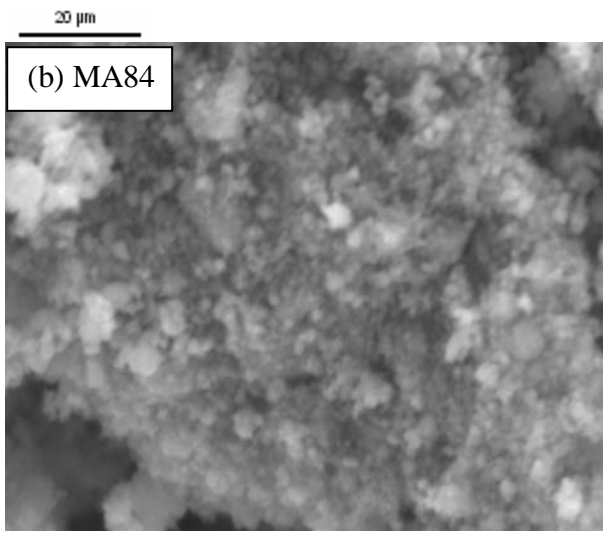
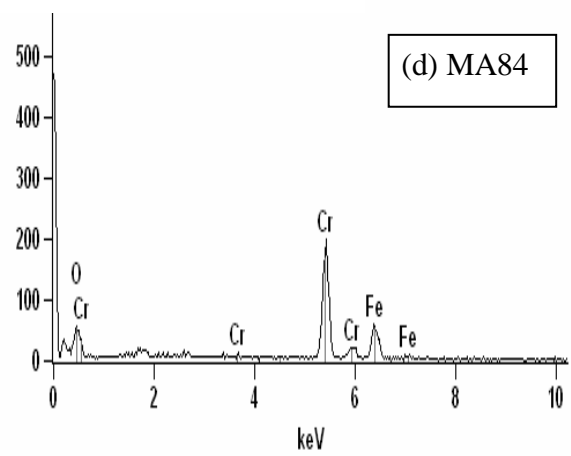

Fig. 2. SEM picture of selected samples MA0 (a), MA84 (b), SN17 (c) and EDAXF spectrum of MA84 sample in (d).

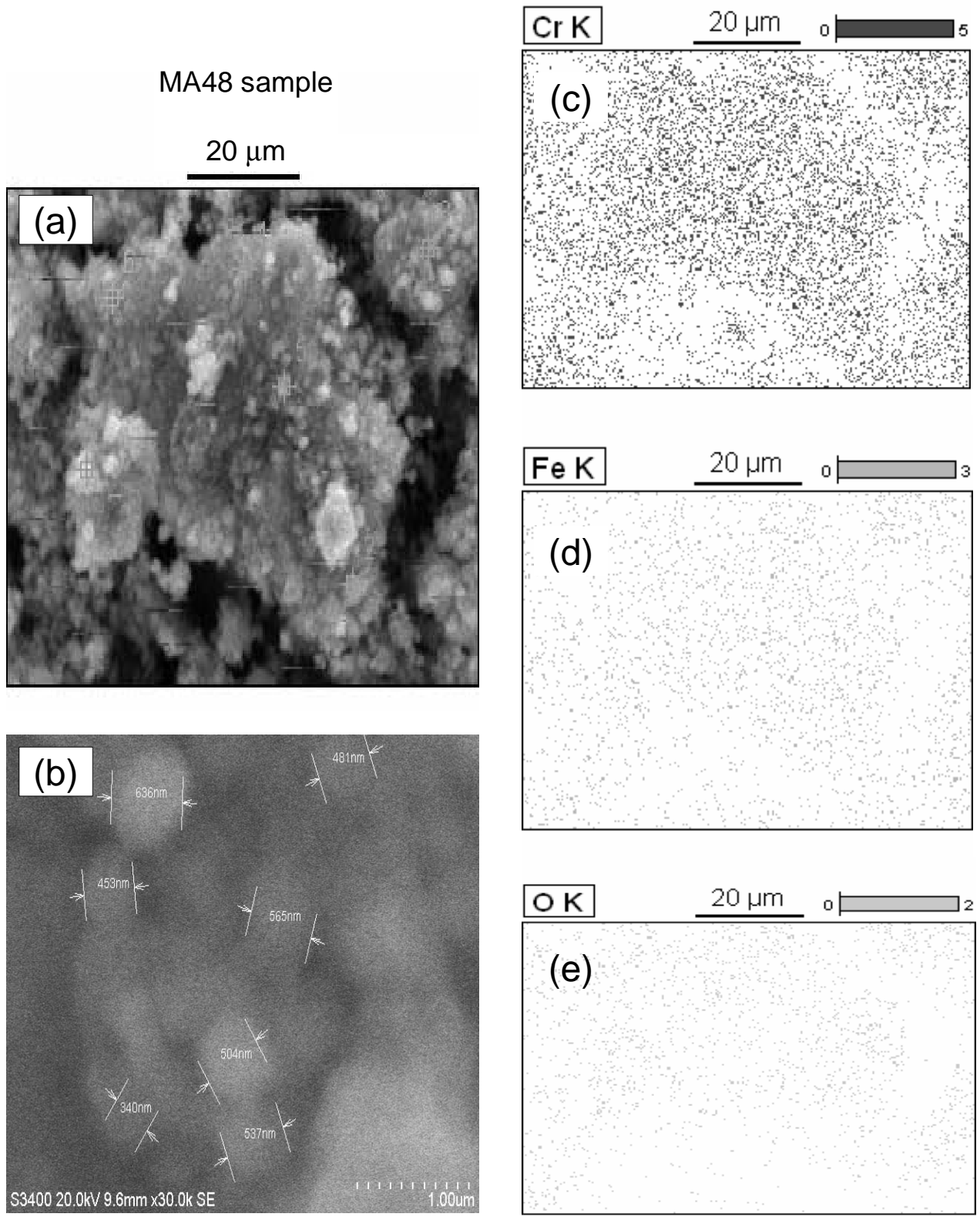

Fig. 3. (Colour online) Selected zone (a) is used to estimate the particle size (b) and mapping of Cr (c), Fe (d) and O (e) atoms in MA48 sample.

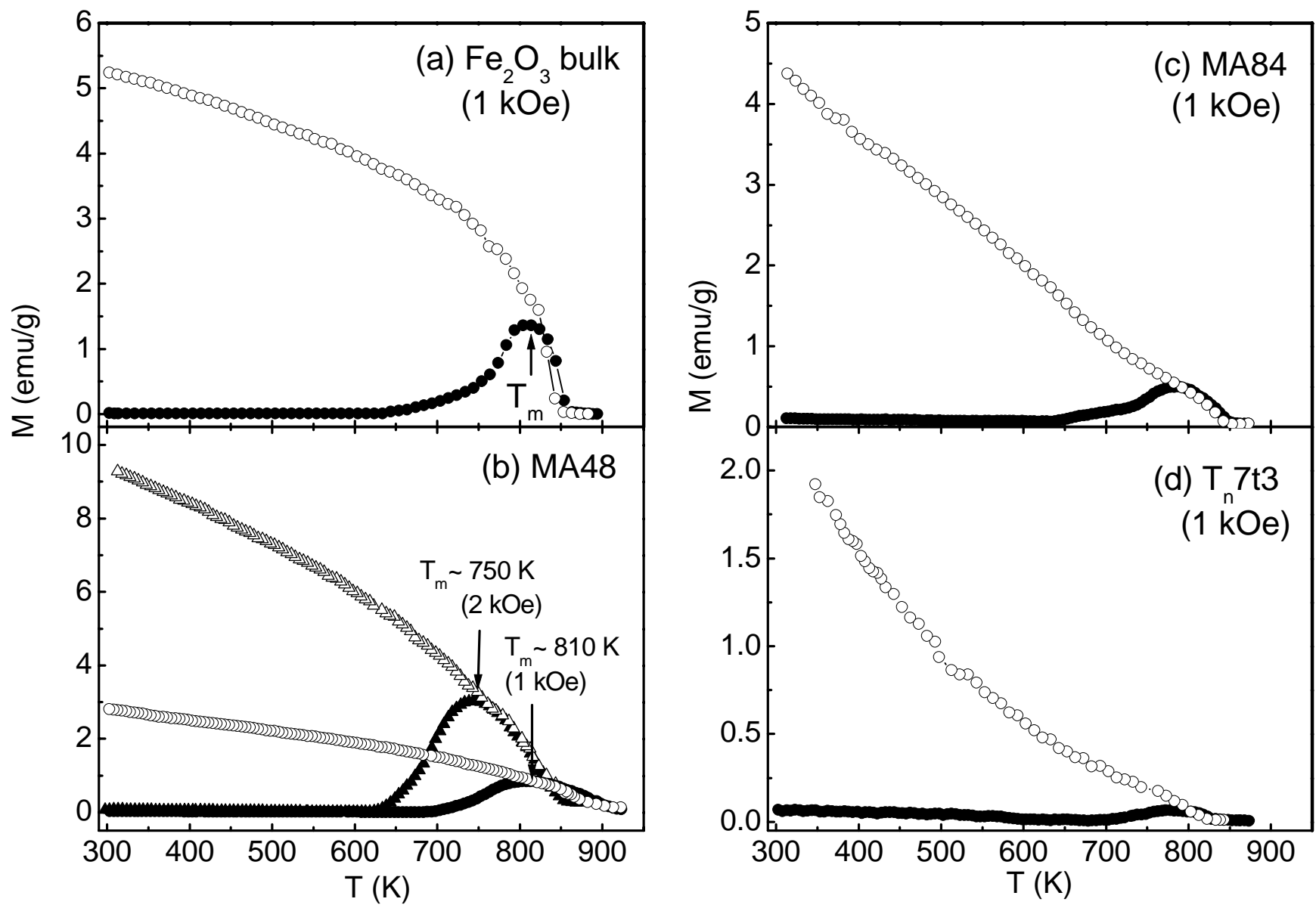

Fig. 4. Temperature dependence of zero field cooled (solid symbol) and field cooled (open symbol) magnetization at 1 kOe for the selected (Fe2O3 bulk, MA48, MA84, SN3) samples. $T_m$ represents the peak temperature of zero field cooled magnetization maximum. Magnetization data at 2 kOe are also shown to show the field dependence of $T_m$.

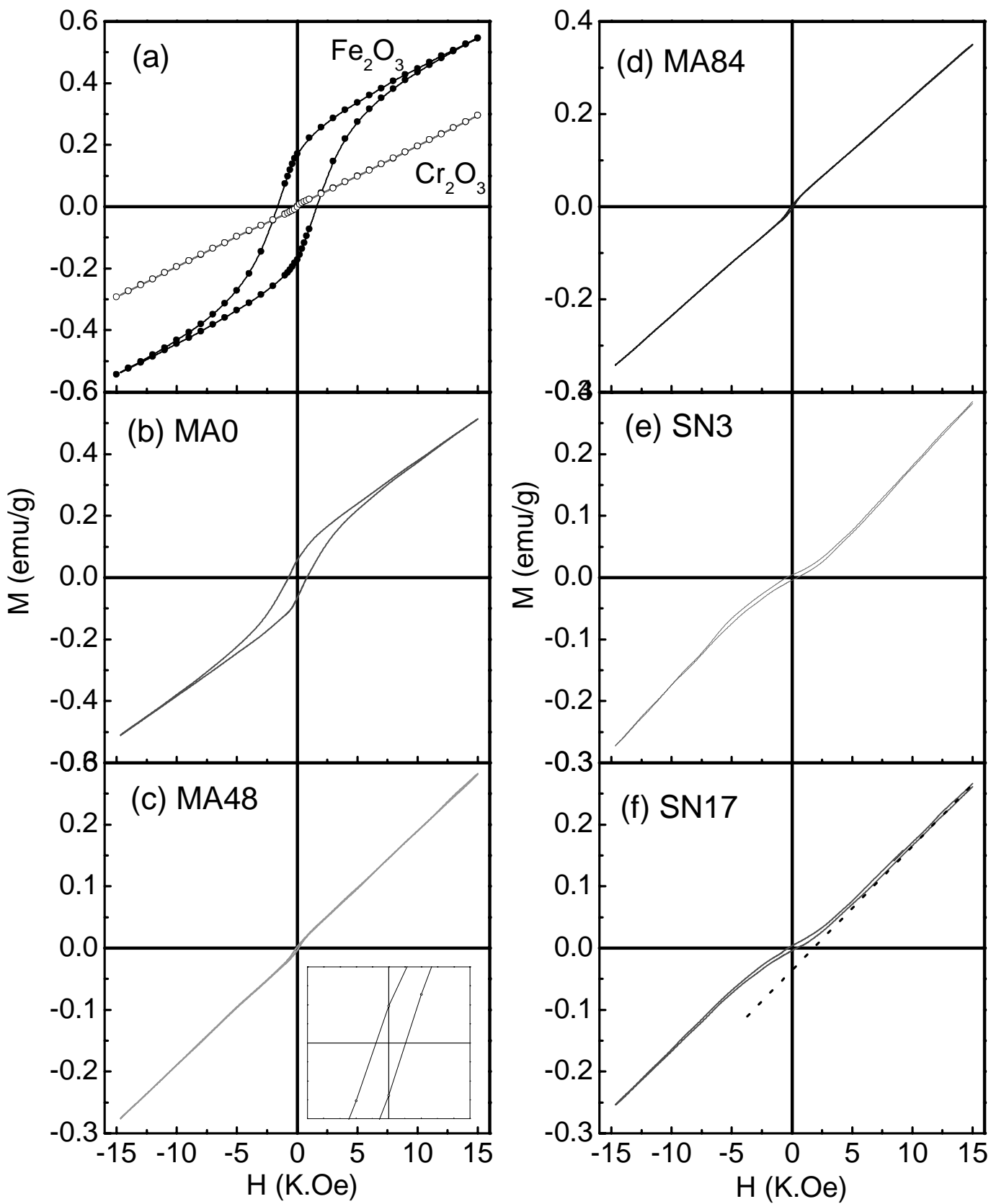

Fig. 5. Variation of magnetization (M) with applied field (H) for different alloyed samples, along with bulk Fe2O3 and Cr2O3 samples. The existence of a small hysteresis loop for MA48 sample is shown in the inset of (c).

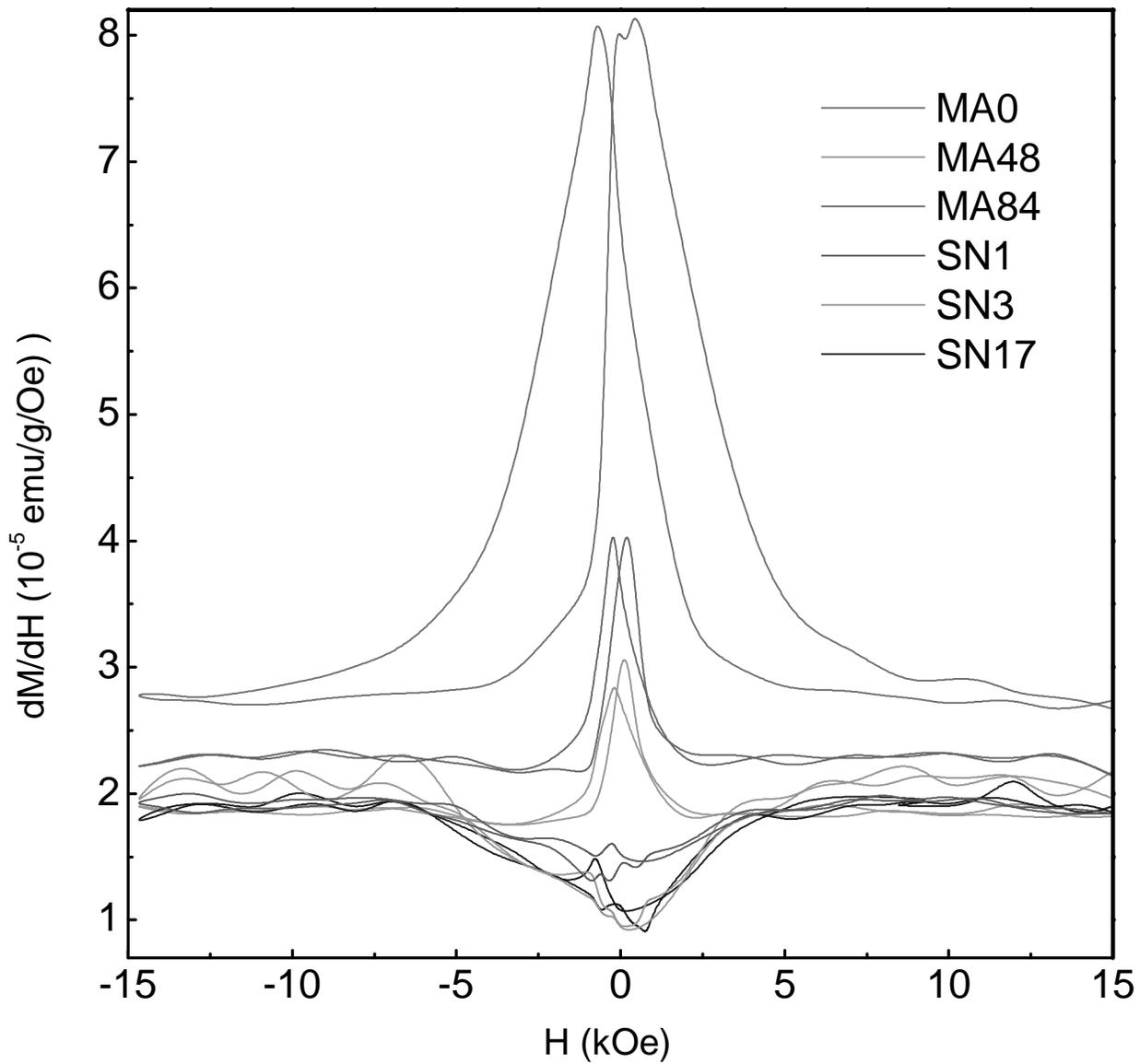

Fig. 6. (Colour online) The field dependence of differential susceptibility (dM/dH) is shown for selected (MA0, MA48, MA84, SN1, SN3, SN17) samples.

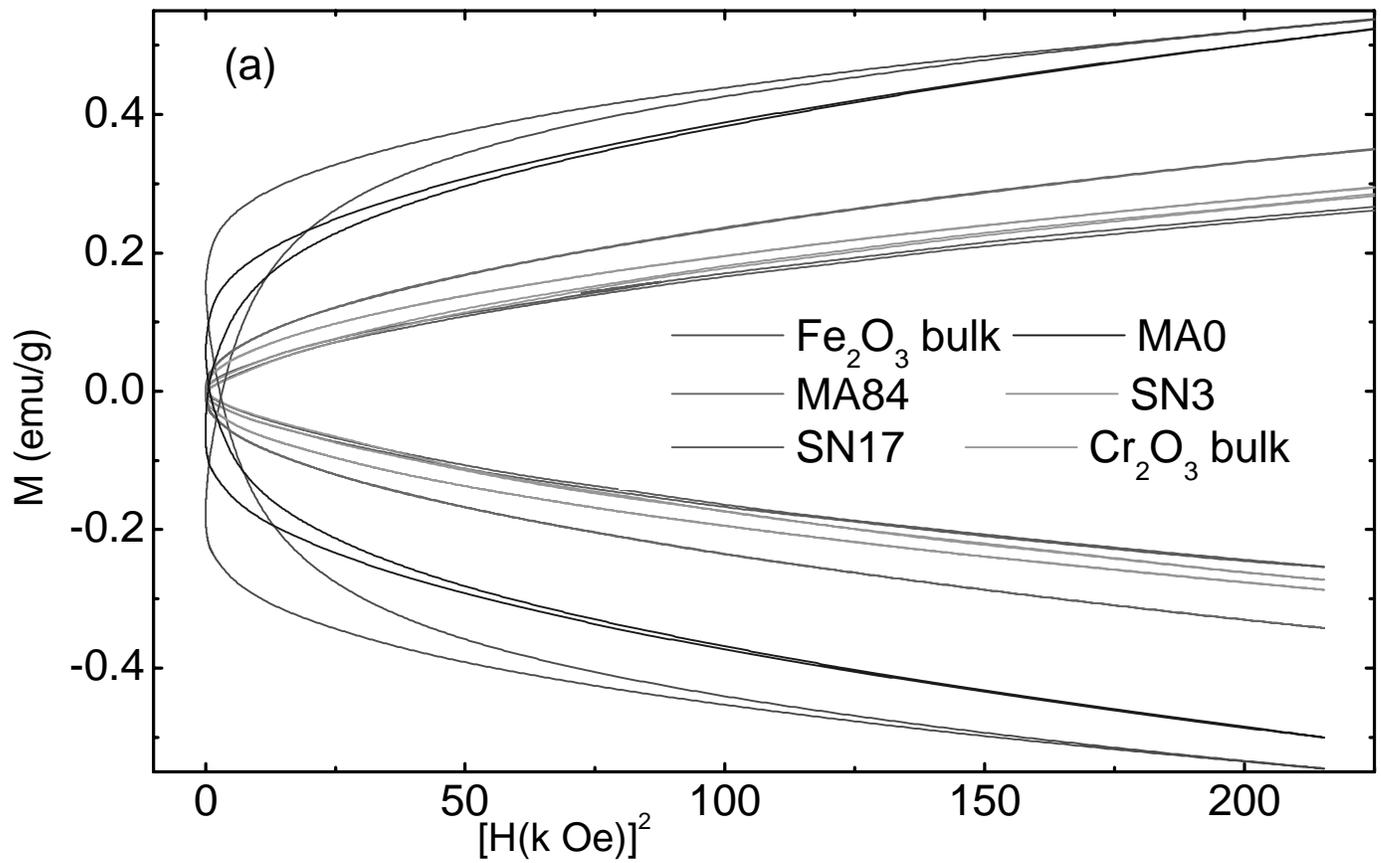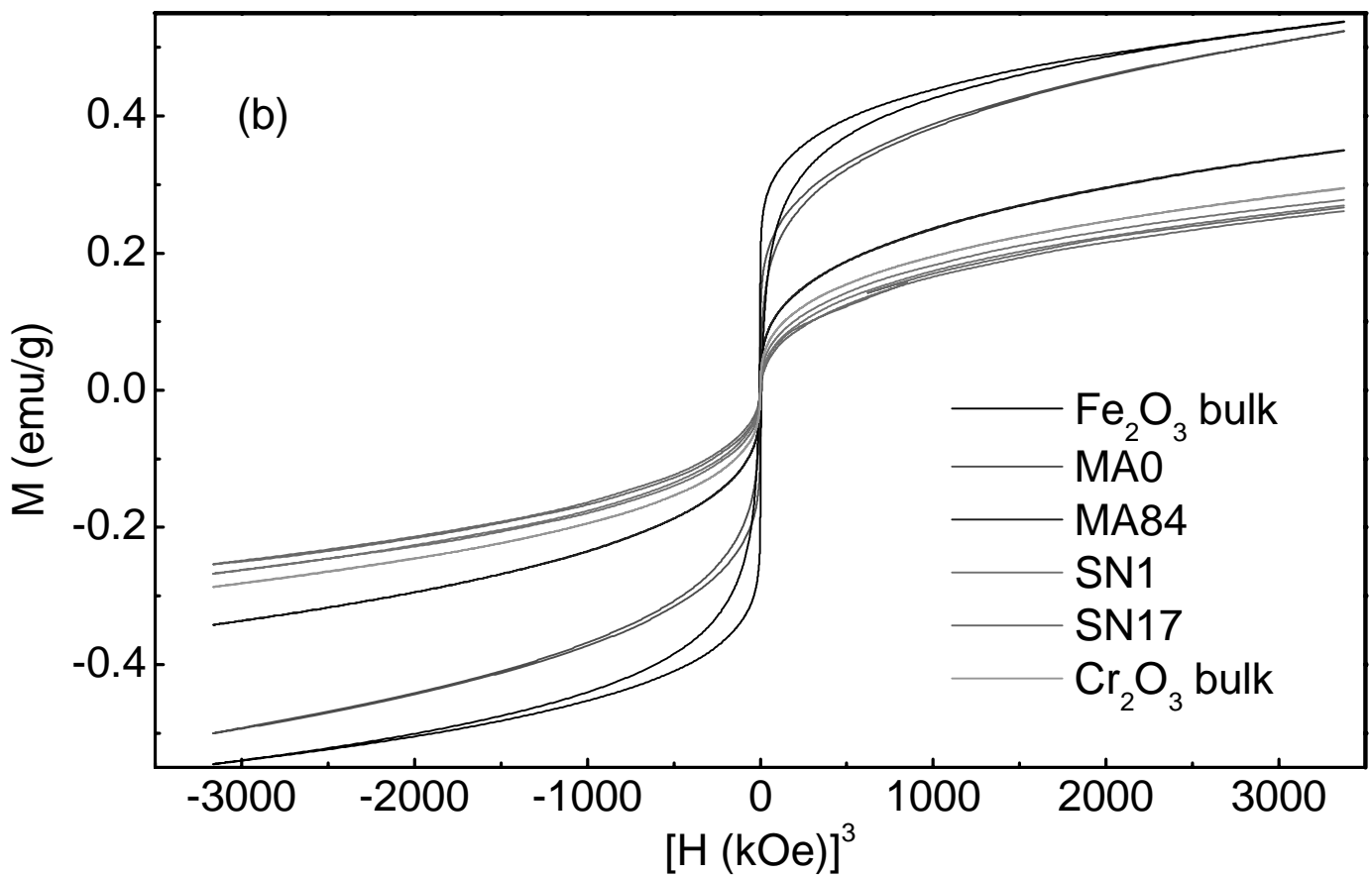

Fig. 7. (Colour online) (a) M vs. H2 plot and (b) M vs. H3 plot for selected samples.

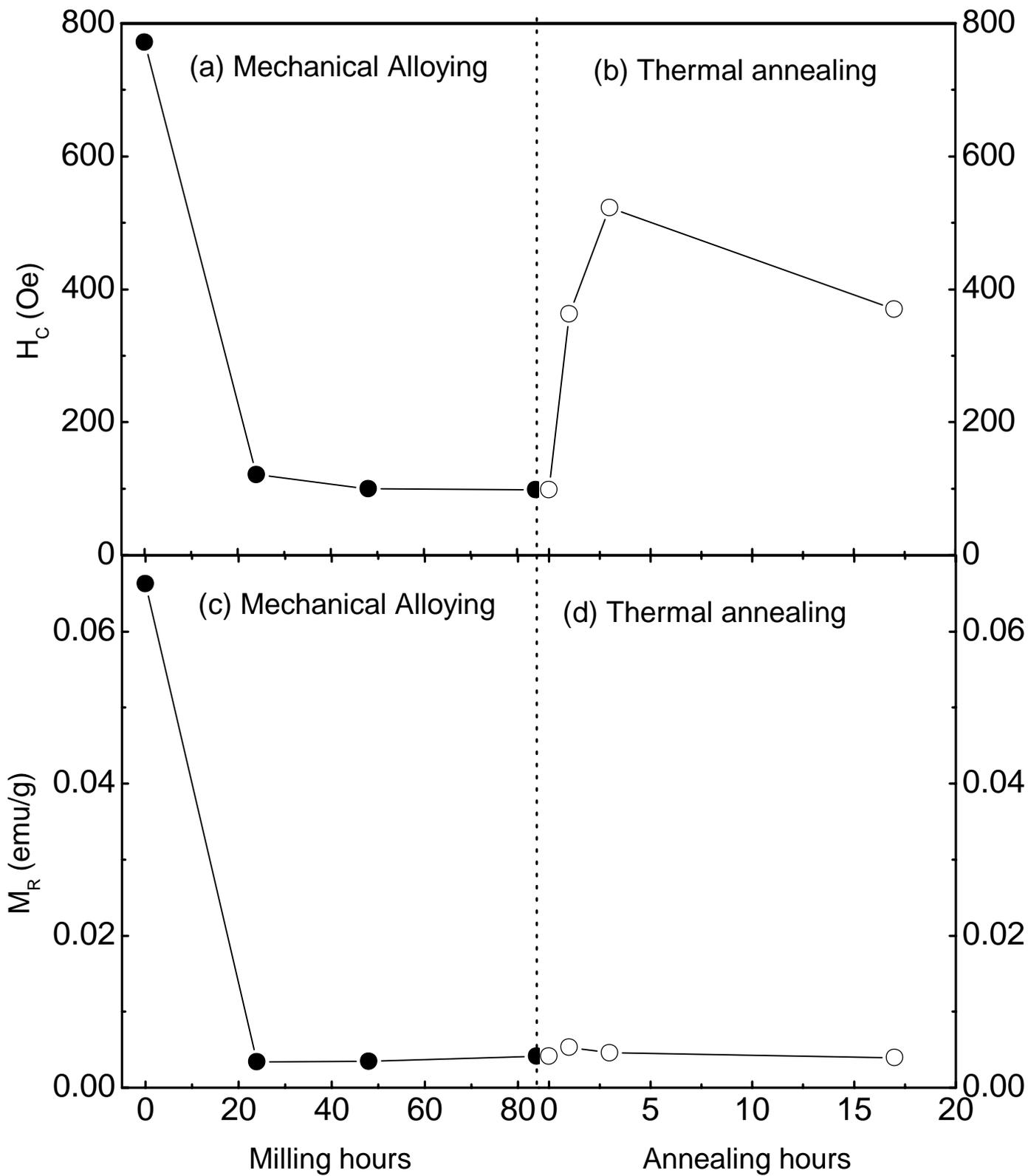

Fig. 8. The variation of coercivity (HC) and remanent magnetization (MR), calculated from hysteresis loop, are shown with the variation of milling time and annealing time.